\documentclass[aps,superscriptaddress,eqsecnum,nofootinbib,preprintnumbers]{revtex4}
\pdfoutput=1
\usepackage{graphicx,epsfig}
\usepackage{amsmath}
\usepackage {amssymb}
\usepackage{placeins}

\newcommand{\be}{\begin{eqnarray}}
\newcommand{\ee}{\end{eqnarray}}
\newcommand{\bea}{\begin{eqnarray}}

\newcommand{\eea}{\end{eqnarray}}

\newcommand{\da}{\dot{a}}
\newcommand{\db}{\dot{b}}

\newcommand{\dda}{\ddot{a}}
\newcommand{\ddb}{\ddot{b}}

\newcommand{\fda}{\frac{\da}{a}}
\newcommand{\fdb}{\frac{\db}{b}}

\newcommand{\fdda}{\frac{\dda}{a}}
\newcommand{\fddb}{\frac{\ddb}{b}}

\newcommand{\fkaa}{\frac{k_a}{a^2}}
\newcommand{\fkbb}{\frac{k_b}{b^2}}

\begin{document}

\title{Reconstruction of Cosmological Evolution in the Presence of Extra Dimensions }

\author{B. C. Georgalas }
\email{vgeorgal@phys.uoa.gr} \affiliation{National and Kapodistrian University of Athens, Department of Physics,
Nuclear and Particle Physics Section, GR–157 71 Athens, Greece}

\author{Stelios Karydas }
\email{stkarydas@mail.ntua.gr} \affiliation{Physics Division,
National Technical University of Athens, 15780 Zografou Campus,
Athens, Greece.}

\author{Eleftherios Papantonopoulos}
\email{lpapa@central.ntua.gr} \affiliation{Physics Division,
National Technical University of Athens, 15780 Zografou Campus,
Athens, Greece.}

\date{\today}

\begin{abstract}
\noindent The model of Physics resulting from a Kaluza-Klein dimensional reduction procedure offers very good dark matter candidates in the form of Light Kaluza-Klein Particles, and thus becomes relevant to Cosmology. In this work we utilize an analytically known, Kasner-type, attractor solution. It is shown that for a plethora of pairs of initial conditions of the usual and extra spatial Hubble parameters, one can recover a late-time cosmological picture, similar to that of the $\Lambda$CDM, in a multidimensional scenario, in the general framework of Universal Extra Dimensions. The phenomenology of fundamental interactions dictates the stabilization of the extra dimensional evolution from a very early epoch in these scenarios. Without an explicit mechanism, this is achieved through particular behaviors of the usual and extra spatial fluids, which have to be motivated by a more fundamental theory.

\end{abstract}
\maketitle

\section{Introduction}

The advances in string theory have put forward the necessity to study models that describe our world in more than four dimensions.
 Then, to recover the four-dimensional spacetime  a dimensional reduction mechanism has to be employed. This mechanism can be realized
by using the Kaluza-Klein (KK) dimensional reduction formalism \cite{Kaluza:1921tu,Klein:1926tv}. In this direction, inspired by string (or M) theory, models were built
in the so-called braneworld scenario \cite{randall}, according to which the Standard Model,
with its matter and gauge interactions, is localized on a
three-dimensional hypersurface (called brane) embedded in a
higher-dimensional spacetime. Gravity propagates in all spacetime
 (called bulk) and thus connects the Standard Model sector with the internal space
dynamics (for a review
 on braneworld dynamics see~\cite{Maartens:2003tw}).

Cosmology in theories with branes
 embedded in  extra dimensions has been the subject of intense investigation.  The most detailed analysis has been done for braneworld models in five-dimensional
   spacetime. The effect of the extra dimension can modify the cosmological evolution, depending on the model,
    both at early  and late times. The
 cosmology of this and other related models
with one transverse to the brane extra dimension (codimension-1
brane models) is well understood.
 In the cosmological generalization of \cite{randall},
 the early-time (high energy limit) cosmological evolution is modified
by the square of the matter density on the brane, while the bulk
leaves its imprints on the brane by the ``dark radiation" term.
 The presence of a bulk cosmological constant
 in \cite{randall} gives conventional
cosmology at late times (low energy limit) (for related reviews see \cite{reviews}).

 In a different approach a model was proposed in \cite{Antoniadis:1990ew}, entailing the existence of large extra dimensions with a size of a few TeV, offering new mechanisms for supersymmetry breaking as one of its primary consequences. A particular case of large extra dimensions is the case referred to as Universal Extra Dimensions (\cite{Appelquist:2000nn}), according to which, the extra dimensions are accessible to all the Standard Model fields. The typical dimensional reduction of the full Lagrangian of any SM particle leads to an infinite tower of KK states that are perceived from a 4-D perspective as massive particles (for related reviews see \cite{Bailin:1987}, \cite{Overduin:2003}).

This setup becomes thus of great interest in cosmological contexts, since it incorporates naturally possible candidates for dark matter. The stability of the Lightest KK Particles (LKPs) could mean they still exist today as thermal relics and if they are not charged and of baryonic nature, they possess all the essential properties of a weakly interacting massive particle (WIMP), (see \cite{Jungman:1995df}, \cite{Servant:2002}, for an introduction to WIMP dark matter).

In the effective 4-D picture of the UED scenario the
fundamental coupling constants vary with the volume of the internal space. However, the strong
cosmological constraints on the allowed variation of these
constants  require the extra space to be not only compactified, but also stabilized, at a time no
later than Big Bang Nucleosynthesis (BBN). Therefore, to produce a viable cosmological model in the UED
scenario, one has to find a dynamical explanation for the stabilization of the extra dimensions in order to reproduce standard cosmology for late times. An effort was made in \cite{Bring:2003-1} to achieve this, and it was shown that while this is possible for the radiation domination era with no explicit mechanism, this is not the case for matter domination (see also \cite{Zhuk:2006kh}, \cite{Eingorn:2010wi}). In particular, by the use of the typical definition of momentum flux for both usual and extra space, it was shown that if KK particles make up a significant part of dark matter, a constraint regarding the equations of state of the usual and extra spatial fluids is obtained, that is in general incompatible with a stabilization constraint coming directly from the field equations. Furthermore, in \cite{Bring:2003-2}, results of a stabilization due to the work of background fields were presented.

In this paper, we work with a quite generic metric that describes a UED scenario, with two separate scale factors for the usual and the extra space. After obtaining the general solution in the phase space of the two Hubble parameters, we show that a special case Kasner-type solution in the presence of matter, that is analytically known, acts as an attractor. Then we reproduce a similar cosmological evolution to that predicted by the $\Lambda$CDM, without imposing any constraint regarding KK dark matter. To do that, one needs to either suppose that KK modes are not a substantial fraction of the dark matter, or abolish the typical connection between momentum flux and pressure for the extra spatial fluid, and instead assume that matter is described in the microscopic level by a more fundamental theory. For example in \cite{Brand:1989}, \cite{Tsetylin:1992}, a case where strings wound around the compactified extra dimensions was proposed, and explored. This results in a negative pressure effect which holds the compactified dimensions in place, impeding their evolution, unlike the usual picture of the non-compactified dimensions where negative pressure is what drives their accelerated expansion (for string gas cosmology reviews see \cite{Brand:2005-2008}, but also \cite{Easther:2003-2004-2005}). In any case, we will recover a picture equivalent to that of the $\Lambda$CDM, in the context of UED, by imposing only (apparent) stabilization constraints quantified by related phenomenology/observations.

This paper is organized as follows: in Section \ref{setupsec} we will present the setup of this scenario, generally following the setup of \cite{Bring:2003-1}. In Section \ref{solutionsec} we will present the solutions of this setup, showing the existence of an attractor solution, to which a huge variety of initial conditions lead, rendering the construction of specific models quite easy. In Sections \ref{constraintsec} and \ref{specificsec} we discuss the constraints that have to be followed, and construct such a model with the sole purpose of recovering the late time results of the $\Lambda$CDM. In Section \ref{scalesec} an interesting case of the equations of state is presented before concluding in Section \ref{conclusion}.

\section{A Homogeneous Universe in $(3+1+n)$-dimensions} \label{setupsec}

We assume that our universe is homogeneous in $(3+1+n)$-dimensions. We also assume that it is not isotropic as a whole but it is isotropic in 3-D and $n$ dimensions separately. This universe  can be described
by the standard Friedmann-Robertson-Walker (FRW) metric, if we allow
for different scale factors in 3-D and $n$ dimensions
\begin{equation}
  ds^2 = -dt^2 + a^2(t) \gamma_{ij} dx^i dx^j + b^2(t) \tilde{\gamma}_{pq}
  dy^p dy^q \,, \label{metric}
\end{equation}
where $\gamma_{ij}$ and $\tilde{\gamma}_{pq}$ are maximally symmetric
metrics in three and $n$ dimensions, respectively. Spatial curvature
is thus parametrized in the usual way by $k_a = -1,0,1$ in ordinary
space, and $k_b = -1,0,1$ in the extra dimensions.

With this choice of metric, the energy-momentum tensor must take the
following form:
\begin{equation}
\label{emtensor}
  T^A_{\phantom{A}B} = \left(
  \begin{array}{ccc}
    -\rho & 0 & 0 \\
    0 & \gamma^i_{\phantom{i} j} p_a & 0 \\
    0 & 0 & \tilde{\gamma}^p_{\phantom{p} q} p_b
  \end{array}
  \right) \,,
\end{equation}
which describes a homogeneous but in general anisotropic perfect fluid
in its rest frame. The pressure in ordinary space  is related
to the energy density by an equation of state $p_a = w_a \rho$, while for the extra space we have $p_b =
w_b \rho$.

The nonzero components of the Einstein field equations in the background metric \eqref{metric} are then given by
\begin{subequations}
\label{allfe}
\begin{eqnarray}
  3\left( \fda \right)^2 + 3\fkaa + 3 n \fda \fdb + \frac{n(n -
    1)}{2}\left[\left( \fdb \right)^2 + \fkbb \right] &=&
  \kappa^2\rho \,, \label{fe00} \\
  2\fdda + \left( \fda \right)^2 + \fkaa + n \fddb + 2n \fda \fdb +
  \frac{n(n - 1)}{2} \left[ \left( \fdb\right)^2 + \fkbb \right] &=&
  -\kappa^2 w_a \rho \,, \label{feij}\\
  3\fdda + 3\left( \fda \right)^2 + 3\fkaa + (n-1)\fddb
  +3(n-1)\fda\fdb + \frac{(n - 1)(n -2)}{2}\left[ \left( \fdb \right)^2 + \fkbb
  \right] &=&-\kappa^2 w_b \rho \,, \label{fepq}
\end{eqnarray}
\end{subequations}
where an overdot denotes differentiation with respect to cosmic time
$t$. From conservation of energy $T^A_{\phantom{A}0;A\phantom{A}} =
0$ we find, furthermore,
\begin{equation}
  \frac{\dot{\rho}}{\rho} = -3(1 + w_a) \fda - n(1 + w_b) \fdb \,.
  \label{econserv}
\end{equation}
For constant equations of state this can be integrated to give
\begin{equation}
  \rho = \rho_i \left( \frac{a}{a_i} \right)^{-3(1 + w_a)} \left(
  \frac{b}{b_i} \right)^{-n(1 + w_b)} \,. \label{rho}
\end{equation}
We will use a subscript $i$ to indicate arbitrary initial values, and a subscript $0$ to indicate today values
throughout.

Introducing the Hubble parameters $H_a=\fda$ and $H_b=\fdb$ for the ordinary and the extra space respectively, equation (\ref{fe00})
becomes
\be
3H_a^2 + 3\fkaa + 3 n H_a H_b + \frac{n(n -
    1)}{2}\left[H_b^2 + \fkbb \right] &=&
  \kappa^2\rho \,. \label{fe010}
    \ee
    Equation (\ref{fe010}) is the Friedmann equation of a homogeneous Universe with energy density $\rho$ in $(3+1+n)$-dimensions with two Hubble parameters. Assuming that the curvature of the three-dimensional space  and also the curvature of the extra dimensional-space are zero, the above relation gives a simple algebraic connection of the Hubble parameter $H_a$ with the Hubble parameter $H_b$ through $\rho$.  The other two equations, (\ref{feij}) and (\ref{fepq}), give the codependent accelerations of the two scale factors. These equations will result to a constraint for the equations of state in order to achieve exact stabilization of the internal space.

    \section{Evolution of the Hubble parameters} \label{solutionsec}

Our purpose is to study how a $(3+1+n)$-dimensional cosmological model can evolve to an effective $(3+1)$-dimensional one. This means that the extra $n$-dimensions have to eventually follow a compactification and stabilization mechanism. It will be shown that a natural way of stabilizing the extra dimensions can be achieved for certain values of the equation  of state parameters  $w_a,$ $w_b$. It is also generally known that the Einstein Field Equations have Kasner-type solutions \cite{Kasner:1921zz}
which are known to act as compactifying mechanisms \cite{Chodos:1979vk}. We will use $k_a=0$, which is the accepted value according to observations. Moreover, we will only consider toroidal compactifications for the extra space, hence also $k_b=0$\footnote{From the equivalent of equation \eqref{fehp02} for $k_b\neq 0$, it can be easily seen that a stabilized extra space is only compatible with $H_a=const.$, which through equation \eqref{fe010} implies $\rho=const.$, that is hard to match with Standard Cosmology.}.

We present here an outline of the process that we follow to obtain the Hubble parameters' evolution of this scenario. First, we impose the ansatz \eqref{ansatz}, between the Hubble parameters, which leads to two Kasner solutions ($K1$, $K2$) that are independent of the Equations of State (EoS) parameters, and one Kasner-type solution ($K3$), which is EoS-dependent and is obtained when the EoS parameters are constant. Then we lift the ansatz \eqref{ansatz}, and by eliminating time from the differential system, we obtain solution \eqref{gensol}, that will be called the ``general solution" of the metric \eqref{metric}. We find that this solution can be written as a product of the K-type particular solutions obtained beforehand, raised to some exponents that are EoS-parameter dependent. We then conclude by studying this solution asymptotically.

     We start by using the relations
	\begin{equation}
		\frac{\ddot{a}}{a}=\dot{H}_a+H_a^2 \text{\,, }\frac{\ddot{b}}{b}=\dot{H}_b+H_b^2~,
	\end{equation}
	and eliminating the energy density by using (\ref{fe010}), we get an equivalent differential system
		\begin{subequations}
\label{fehp}
\begin{eqnarray}
  \dot{H}_a&=&\frac{3\bigl[(n-1)w_a-nw_b-n-1\bigr]}{2+n}H_a^2+\frac{n\bigl[ (n-1)(3w_a-1)-3nw_b\bigr]}{2+n}H_a H_b\nonumber \\
  &+&\frac{n (n-1)\bigl[1+(n-1)w_a-nw_b \bigr]}{2(2+n)}H_b^2 \,, \label{fehp01} \\
  \dot{H}_b&=&\frac{3\left(2 w_b-3 w_a+1\right)}{2+n}H_a^2-\frac{3\left( 3nw_a-2nw_b+2\right)}{2+n}H_a H_b \nonumber \\
  &-&\frac{n\bigl[5+n+3(n-1)w_a-2(n-1)w_b\bigr]}{2(2+n)}H_b^2~.  \label{fehp02}
\end{eqnarray}
\end{subequations}

This system of the two Hubble parameters depends only on the EoS parameters $w_a$ and $w_b$. Looking for particular solutions, we impose the ansatz
\begin{equation} \label{ansatz}
H_b(t)=c_i H_a(t)
\end{equation}
Substituting it in \eqref{fehp}, we get two equations for $\dot{H_a}$. Demanding that these equations have the same solution gives 3 possible values for $c_i$ when $n\geq 2$:

\begin{equation} \label{ckas}
\underbrace{c_1=\frac{6}{-3n-\sqrt{3n(2+n)}}}_\text{K1}   \qquad   \underbrace{c_2=\frac{6}{-3n+\sqrt{3n(2+n)}}}_\text{K2}   \qquad   \underbrace{c_3=\frac{1-3 w_a+2 w_b}{1+(n-1) w_a-n w_b}}_\text{K3}
\end{equation}
while for $n=1$
\begin{equation} \label{ckasn1}
\underbrace{c_1=-1}_\text{K1}   \qquad   \underbrace{c_3=\frac{-1+3w_a-2w_b}{-1+w_b}}_\text{K3}
\end{equation}
For $n\geq 2$ these are the two Kasner solutions $K1$ and $K2$:
\begin{equation} \label{kasner1}
\left.\begin{aligned}
H_a(t)&=\frac{H_a(0)(n-1)}{n-1+[\sqrt{3n(2+n)}-3]H_a(0)t} \\
   H_b(t)&=-\frac{6H_a(0)}{3n+\sqrt{3n(2+n)}+[3n+3\sqrt{3n(2+n)}]H_a(0)t}
\end{aligned}
\right\}
\qquad \text{K1}
\end{equation}

\begin{equation} \label{kasner2}
\left.\begin{aligned}
H_a(t)&=\frac{H_a(0)(n-1)}{n-1-[\sqrt{3n(2+n)}+3]H_a(0)t}\\
   H_b(t)&=\frac{6H_a(0)}{-3n+\sqrt{3n(2+n)}+[-3n+3\sqrt{3n(2+n)}]H_a(0)t}
\end{aligned}
\right\}
\qquad \text{K2}
\end{equation}
while for $n=1$ there is only one Kasner solution (the second one essentially reduces to the trivial $H_a=0$, as will be shown later):
\begin{equation} \label{kasner1n1}
\left.\begin{aligned}
H_a(t)&=\frac{H_a(0)}{1+2H_a(0)t} \\
   H_b(t)&=-\frac{H_a(0)}{1+2H_a(0)t}
\end{aligned}
\right\}
\qquad \text{K1 for $n=1$}
\end{equation}	
It seems peculiar that \eqref{fehp}, which has explicit dependence on the EoS parameters, has as solutions expressions that do not depend on them, however we note that originally the Kasner solutions were vacuum solutions of the Einstein equations \cite{Kasner:1921zz}. So by returning to the system \eqref{allfe}, from which system \eqref{fehp} is produced through algebraic manipulation, and using $\rho=0$, we see that all $w$-dependences are switched off. In
\cite{Chodos:1979vk} it was discussed how Kasner solutions can be generalized in the presence of matter.

When the $w$ parameters are constant, the system \eqref{fehp} has another Kasner-type solution (hinted at in \cite{Gu:2002}), which will be called $K3$ solution:

\begin{equation} \label{k3sol}
\left.\begin{aligned}
H_a(t)&=&\frac{2[1+(n-1)w_a-nw_b]H_a(0)}{2+2(n-1)w_a-2nw_b+[3-3w_a^2+n(1+3w_a^2-6w_a w_b+2w_b^2)]H_a(0)t}\\
   H_b(t)&=&\frac{2(1-3w_a+2w_b)H_a(0)}{2+2(n-1)w_a-2nw_b+[3-3w_a^2+n(1+3w_a^2-6w_a w_b+2w_b^2)]H_a(0)t}
\end{aligned}
\right\}
\qquad \text{K3}
\end{equation}

	All these particular solutions obviously have constant $H_b/H_a$ ratios throughout their evolution. Naturally, one would look for the Hubble parameters' evolution, corresponding to the metric \eqref{metric}, without imposing the ansatz \eqref{ansatz}. As we will show in a moment, this evolution is actually made up as a product of the particular solutions corresponding to the ansatz \eqref{ansatz}, with $c_i$ given by \eqref{ckas}.
	
	Because of their form $(1/t)$, one understands that these solutions have a singularity and some interesting properties. The Kasner solutions ($K1$, $K2$) have a constant deceleration parameter throughout. For $n=1$ we have $q_{K1}=1$ while for $n\geq 2$:
	\[
	q=-1-\frac{\dot{H_a}}{H_a^2}=
	\begin{cases}
	\frac{2+n-\sqrt{3 n (2+n)}}{1-n}>0 \text{ }\forall \text{ }n\text{ } \geq\text{ } 2\qquad (K1) \\
	\frac{2+n+\sqrt{3 n (2+n)}}{1-n}<0 \text{ }\forall \text{ }n \text{ }\geq \text{ }2\qquad (K2)
	\end{cases}
	\]
In the case of $K3$ however, the sign of q depends also on the values of the $w$ parameters:
\[
	q=\frac{1+n+(2-2n)w_a-6nw_a w_b+2 n w_b+(3n-3)w_a^2+2 n w_b^2}{2+2(n-1)w_a-2 n w_b}
\]
Moreover, for a positive value of $H_a$ for $t=0$, $K1$ solution has its singularity for $t<0$ and $K2$ for $t>0$, while $K3$'s singularity again depends on the $w$ parameters. Finally, it is notable that for the $K1$, $K2$ solutions a contraction of the extra space ($H_b<0$) guarantees the growing ($H_a>0$) of the 3-d space and vice versa, while that is not necessarily true for $K3$ because of the $w$ parameters.
	
	Moving on, to obtain the ``general solution" that the Friedmann equations yield for the metric \eqref{metric}, without imposing any ansatzes of the form of \eqref{ansatz}, we eliminate the time parameter in \eqref{fehp}, thus passing to a single differential equation, that is always integrable when the EoS parameters, $w$, are constant:
	\begin{align}
	\frac{dH_a}{dH_b}&= \frac{6\bigl((n-1)w_a+n w_b+n+1\bigr)H_a^2+2n\bigl(n-1-3(n-1)w_a+3nw_b\bigr)H_a H_b}{6(3w_a-2w_b-1)H_a^2+6\bigl(n(3w_a-2w_b)+2\bigr)H_a H_b+n\bigl[5+n+(n-1)(3w_a-2w_b)\bigr]H_b^2} \nonumber \\
	&- \frac{n(n-1)\bigl(1+(n-1)w_a-n w_b\bigr)H_b^2}{6(3w_a-2w_b-1)H_a^2+6\bigl(n(3w_a-2w_b)+2\bigr)H_a H_b+n\bigl[5+n+(n-1)(3w_a-2w_b)\bigr]H_b^2}
\end{align}
		Integrating this equation, we get a solution that can, with some algebraic manipulation, be written as a product of the above particular solutions in the form:
		\begin{align} \label{gensol}
	const.=&\Bigl\lvert\underbrace{H_b}_\text{$H_b$ part}\Bigr\rvert^{\sqrt{2+n} \bigl[ 3 (w_a-1)^2+n \left(1-3 w_a^2+6 w_a w_b-2 w_b (1+w_b)\right)\bigr]}\cdot \nonumber \\
	&\Bigl\lvert  \underbrace{\frac{H_a}{H_b}+\frac{3n+\sqrt{3n}\sqrt{2+n}}{6}}_\text{$K1$ part}  \Bigr\rvert^{\sqrt{2+n}(3+n-3w_a-nw_b)+\sqrt{3n}(2+n)(w_a-w_b)}\cdot \nonumber \\
	&\Bigl\lvert  \underbrace{\frac{H_a}{H_b}+\frac{3n-\sqrt{3n}\sqrt{2+n}}{6}}_\text{$K2$ part}  \Bigr\rvert^{\sqrt{2+n}(3+n-3w_a-nw_b)-\sqrt{3n}(2+n)(w_a-w_b)}\cdot \nonumber \\
	&\Bigl\lvert\underbrace{\frac{(n-1)w_a-nw_b+1}{3w_a-2w_b-1}+\frac{H_a}{H_b}}_\text{$K3$ part}\Bigr\rvert^{-\sqrt{2+n}\bigl(3-3w_a^2+n(1+3w_a^2-6w_aw_b+2w_b^2)\bigr)}
\end{align}

This form of the solution of the Hubble parameters that correspond to the metric \eqref{metric}, as a product of the special solutions of the system, ought to be expected due to the similarity of the resulting differential equation, with the Darboux equation (see \cite{Ince}, \textsection 2.21). It should be noted that during the derivation of this solution, $K1$, $K2$, and $K3$ in the form $H_b-c_i H_a=0$, become forbidden constraints, since they appear in denominators of partial fractions. Hence, their corresponding curves in the space of $H_a$, $H_b(H_a)$ (which will be called phase space from now on), will appear as \textit{limiting curves} of every other possible phase curve.  We will limit our study only to solutions that are cosmologically relevant (i.e. excluding solutions with contracting 3-space: $H_a<0$), but a more general picture can be found in the Appendix, in the form of flow diagrams. We also note that the Kasner-type curves will obviously be straight lines in this space, which in the case of $K1$ and $K2$ depend only on n (see eq. \eqref{ckas}), and thus are the same regardless of the $w$ parameters of the model, while for $K3$ the ratio $H_b/H_a$ depends also on the $w$ parameters.

From \eqref{gensol} we can study the solution asymptotically, distinguishing two main cases. The first one  $H_a$, $\lvert H_b \rvert\rightarrow \infty$ uniformly, with $H_a>0$ could correspond to the behavior of a universe close to a singularity, while the case $H_a$, $\lvert H_b \rvert\rightarrow 0$ describes the asymptotic behavior of a universe that goes towards an ``equilibrium" state and is the only case that we will need to match this setup with the standard evolution\footnote{It can be seen directly from \eqref{gensol} that asymptotic cases like $H_a/H_b\not\rightarrow c$ are not possible. For example in a case where asymptotically $H_a \gg H_b$, $H_a/H_b\not\rightarrow c$, if we ignore the non important constants, we would end up with an equation of the form $const.=\rvert H_b\lvert^{...} \lvert \frac{H_a}{H_b}\rvert^{...}$, giving $\rvert H_b\lvert H_a^x=const.$. But by substituting this constraint in \eqref{fehp}, we see that a non-trivial solution of this type is not possible. In this manner one can see that asymptotically it is only possible for one of the $K1$, $K2$, $K3$ solutions to end up attracting the phase curve of any other solution.}.

Because of the form of the exponents to which the $K1$, $K2$, $K3$ solutions are raised in the general solution \eqref{gensol}, the above asymptotic behaviors can be reached differently, depending on the powers' signs, as well as on the position of the initial values $\bigl(H_a(0),H_b(0)\bigr)$ with respect to the $K1$, $K2$, $K3$ curves, essentially determining towards which of the $(H_a,H_b)$ pairs: $(0,0)$, $(\pm \infty,\pm \infty)$ (and $(0,\pm \infty)$ if $n=1$) the solution goes asymptotically\footnote{This is only incidentally true in the cases that interest us. To be more precise the deciding factors for the attractors are the combinations of the signs of the factors in \eqref{fehp} along with the position of the initial conditions compared to the $K1$, $K2$, $K3$ curves.}.

To see this more clearly, we will work without any loss of generality with the solution for $n=1$, from where we can incidentally see that the $K2$ part reduces to the trivial solution $H_a=0$:

\begin{align} \label{gensoln1}
	const.=&\bigl\lvert\underbrace{H_b}_\text{$H_b$ part}\bigr\rvert^{2 \sqrt{3} (3 w_a-w_b-2) (w_b-1)}\cdot
	\Bigl\lvert  \underbrace{\frac{H_a}{H_b}+1}_\text{$K1$ part}  \Bigr\rvert^{\sqrt{3}(4-3w_a-w_b)+3\sqrt{3}(w_a-w_b)}\cdot \nonumber \\
	&\Bigl\lvert  \underbrace{\frac{H_a}{H_b}}_\text{$K2$ part}  \Bigr\rvert^{\sqrt{3}(4-3w_a-w_b)-3\sqrt{3}(w_a-w_b)}\cdot
	\Bigl\lvert\underbrace{\frac{1-w_b}{3w_a-2w_b-1}+\frac{H_a}{H_b}}_\text{$K3$ part}\Bigr\rvert^{-\sqrt{3}(4-6 w_a w_b+2 w_b^2)}
\end{align}

\begin{figure}[h]
\centering
\includegraphics[width=0.49\textwidth]{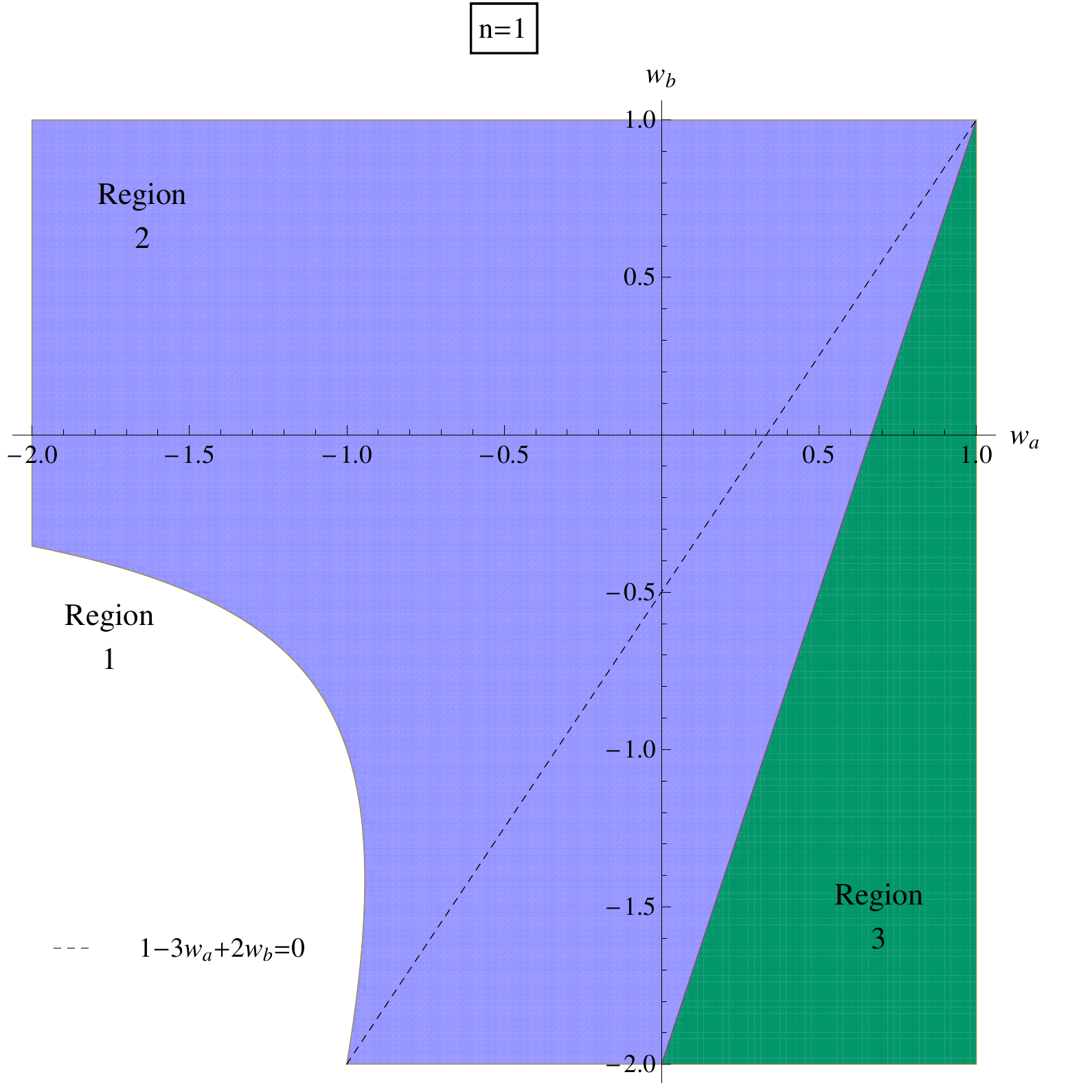}%

\caption{For $n=1$, the term that only contains $H_b$ in \eqref{gensoln1} is raised to an exponent that is positive everywhere except region 3. The exponent to which the $K3$ part is raised is negative in regions 2 and 3. The $K1$ part is raised to a positive power everywhere, while the $K2$ part is positive in regions 1 and 2. So for example in the case where $\frac{H_a}{H_b}\rightarrow const$ with $H_a$, $H_b\rightarrow 0$ the general solution can be consistent in region 2 where the $H_b$ part goes to 0, but can be canceled out by the $K3$ part that goes to infinity. When $H_a$, $H_b \rightarrow \infty$, the roles are reversed, and the $K1$ part is canceling out the $H_b$ part in region 2. Similar conclusions can be reached for any $n$, although the various regions are different.}
\label{consgraphs}
\end{figure}

We show the regions where the exponents of \eqref{gensoln1} have specific signs as functions of the $w$ parameters in Figure \ref{consgraphs}, noting that the region most relevant to a cosmological model is region 2. This is because it includes the constraint $1-3w_a+2w_b=0$, which as shown in Section \ref{constraintsec}, is a necessary condition in order to keep many results of Standard Cosmology intact.

To illustrate the above, we will work as an example in the case $H_a/H_b\rightarrow const$, with $H_a$, $H_b\rightarrow 0$, where we see that if we have chosen $w$ parameters in region 2 of Figure \ref{consgraphs}, the $H_b$ part of \eqref{gensoln1} will go to 0, since it is raised to a positive exponent. Since \eqref{gensoln1} is made up as a product of various factors, at least one of them needs to go to infinity, to nullify the $H_b$ part going to 0, and thus be consistent with the constant value of \eqref{gensoln1} on the $l.h.s$. Assuming we have chosen appropriate initial values\footnote{Meaning initial values that correspond to a phase curve that is limited by the $K1$ and $K3$ curves in this example.}, that can only be achieved asymptotically if $H_a/H_b\rightarrow 1/c_3$, thus making the $K3$ part's base go to zero raised to a negative exponent, so in total going to infinity. If on the other hand $H_a/H_b\rightarrow const$ with $H_a$, $H_b\rightarrow \infty$, the only way to have consistency in \eqref{gensoln1} with $w$ parameters in region 2, is if the $K1$ part goes to 0 (so $ H_a/H_b\rightarrow 1/c_1)$, nullifying the $H_b$ part that now goes to infinity.
	
\begin{figure}[h]
	
	\centering
		\includegraphics[width=0.85\textwidth]{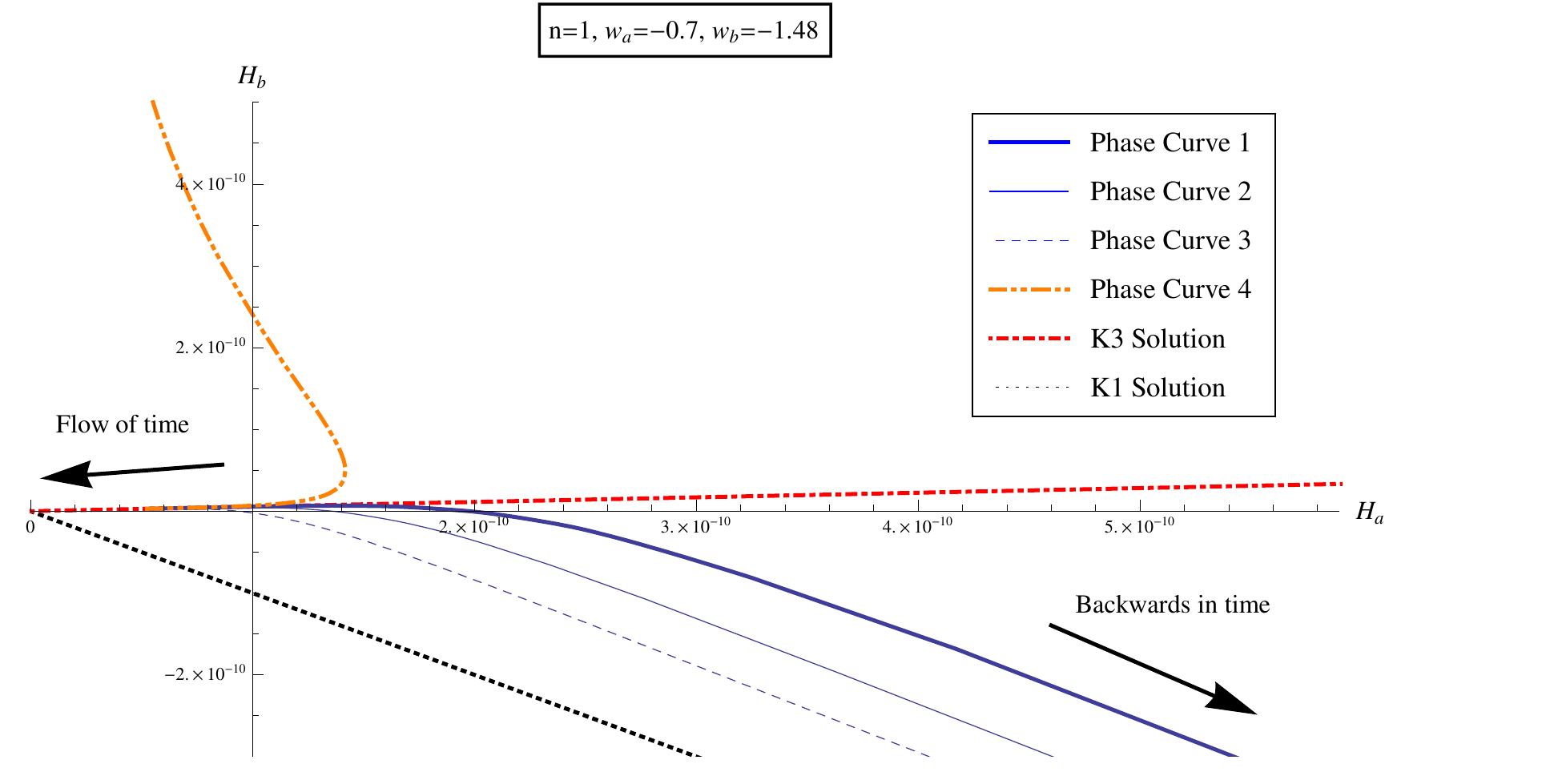}
		\caption{The phase curves for 4 different choices of today values of $H_a$, $H_b$. For curves $1-3$, with today conditions increasingly further from the $K3$ solution, the general solution is dominated by the $K3$ part for a smaller region (that generally translates to a smaller time period). Curve \#4 corresponds to a today ratio $H_b/H_a$ that is bigger than that of the $K3$ solution, hence it corresponds to a different family than the first 3 curves. Going backwards in time one can see that the $K1$ solution starts to dominate for curves $1-3$, while for the fourth curve the part that dominates is $K2$, which for $n=1$ degenerates to the curve $H_a=0$.}
		\label{phasegraph1}
\end{figure}

We can use this mathematical result to construct cosmological models with the desired properties, since region 2 of Figure \ref{consgraphs} contains, in terms of the w parameters, all the cosmologically relevant values that we will need, giving us a great freedom: For any initial values that are contained between the $K1$ and $K3$ curves, and $w$ parameters in region 2, we know exactly the asymptotic behavior of the corresponding solution, which will converge to the $K3$ special solution as $H_a$, $H_b\rightarrow 0$. So, by making the $K3$ solution have some desired properties by means of fixing the $w$ parameters (for example $q<0$ and $\lvert H_b\rvert \ll H_a$), we actually force the general solution \eqref{gensol} to eventually behave like that as well.
We illustrate what was discussed here in Figure \ref{phasegraph1}, where for this reason we have chosen a specific pair of $w$ parameters and construct numerically the phase curves for 4 different choices of initial conditions, showing the behaviors of their phase curves as compared to the phase curves of the Kasner-type solutions.

Concluding this section, it is interesting to note that this asymptotically attracting behavior of the Kasner-type solutions essentially translates to an attractor for the energy density, $\rho$, through equation \eqref{fe010} (for $k_a=k_b=0$). The energy density of these types of scenarios can and will ultimately be attracted to either empty universe scenarios ($\rho=0$) through $K1$ and $K2$ solutions, or to the value predicted by the $K3$ solution:
\begin{equation}
k^2\rho=-\frac{2 (2+n) \bigl[-3(w_a-1)^2+n \bigl(3 w_a^2-6 w_a w_b+2 w_b (1+w_b)-1\bigr)\bigr] H_a(0)^2}{\bigl[2+2 (n-1) w_a-2 n w_b+t \bigl(3-3 w_a^2+n (1+3 w_a^2-6 w_a w_b+2 w_b^2)\bigr) H_a(0)\bigr]^2}.
\end{equation}
\FloatBarrier

		\section{Constraints to obtain a Cosmologically Viable Model} \label{constraintsec}

As we have argued in Section \ref{solutionsec}, any random pair of initial conditions in the space of $H_a$, $H_b(H_a)$ will correspond to a solution that converges to one of the three particular solutions corresponding to \eqref{ansatz} and \eqref{ckas}. This is not particularly interesting from a realistic cosmological point of view in the case of the $K1$ and $K2$ solutions, which always give a specific, EoS-independent, evolution for $H_a$, $H_b$. However, this is not the case for the $K3$ solution, since, through its dependence on the EoS parameters, it can be manipulated to produce certain desirable behaviors. And as soon as the $K3$ solution is fixed to follow some specific constraints, the same will be true for a large variety of solutions corresponding to random initial conditions, that are of the form of the ``general" solution \eqref{gensol}, of the metric \eqref{metric}.

Naturally, the next step is to quantify the constraints needed to be followed to have stabilized extra dimensions, in order to produce a plausible cosmology. If we then translate them into what the behaviors of the EoS parameters need to be, for the $K3$ solution (and subsequently any other solution of the form \eqref{gensol} that converges to it), we will have created a vast family of initial conditions that ultimately lead to a viable cosmology.

Firstly, it is easy to obtain a constraint for an exact stabilization regarding the $w$ parameters(\cite{Bring:2003-1}). By inspecting the equation of the extra dimensional Hubble parameter, \eqref{fehp02}, switching off all terms that contain $H_b$ leaves as the only non trivial solution the equation\footnote{This particular combination of the EoS parameters is affected solely by the dimensionality of the usual space, and not that of the extra space. If for example the scale factor $a(t)$ corresponded to $m$ instead of 3 dimensions, this constraint would be $1-m w_a+(m-1)w_b=0$. The $c_i$ would be different too, for example $c_3=\frac{1-m w_a+(m-1)w_b}{1-(n-1)w_a-n w_b}$}:
		\begin{equation} \label{stabcons}
			1-3 w_a+2 w_b=0~.
		\end{equation}

		For this work, however, we will not impose this particular constraint, but instead we will allow for a slow evolution of the extra dimensions, which, as one would expect, leads to a looser version of \eqref{stabcons}. To see that, one needs only to observe that \eqref{stabcons} is actually a special case of the $K3$ solution - that in which we demand that $c_3=0$. In the phase space $H_a$, $H_b(H_a)$, that corresponds to the $H_b=0$ axis. So to enforce a looser version of \eqref{stabcons} we would simply need to demand that $w_a$, $w_b$ be such as to produce a $K3$ curve whose ratio $H_b/H_a$ is very small.

In particular, to build a viable model, two specific constraints will be used to quantify the apparent to an effective
3-dimensional observer stabilization. The first one is:
		\begin{equation} \label{cons1}
			\lvert H_b^{(0)}\rvert <\frac{1}{10n}H_a^{(0)}~,
		\end{equation}
which is a constraint derived in \cite{Cline:2003}, by comparing the experimental/observational results for $\frac{\dot{G_N}}{G_N}$ with the accepted value of the Hubble parameter $H_0$, and using the fact that $G_N \propto b^{-n}$. The second constraint is:
		\begin{equation} \label{cons2}
			\frac{\lvert b_{BBN}-b_{today}\rvert}{b}\approx 1\%~.
		\end{equation}
where $b_{BBN}$ refers to the value of the scale factor during the BBN era. Constraint \eqref{cons2} can be inferred from various works that take into account a variety of tests, (for example constraints on element abundances that can be used to check the electroweak coupling for redshifts referring as far back as BBN, to more recent redshifts from events like the Oklo natural reactor, see \cite{Uzan:2003} and references therein).

Stricter constraints can of course be applied and still produce solutions. However, much stricter constraints will effectively lead to an exact stabilization, which as already mentioned, is merely a special case of the $K3$ solution.

\begin{figure}[h]
	
	\centering
		\includegraphics[width=0.5\textwidth]{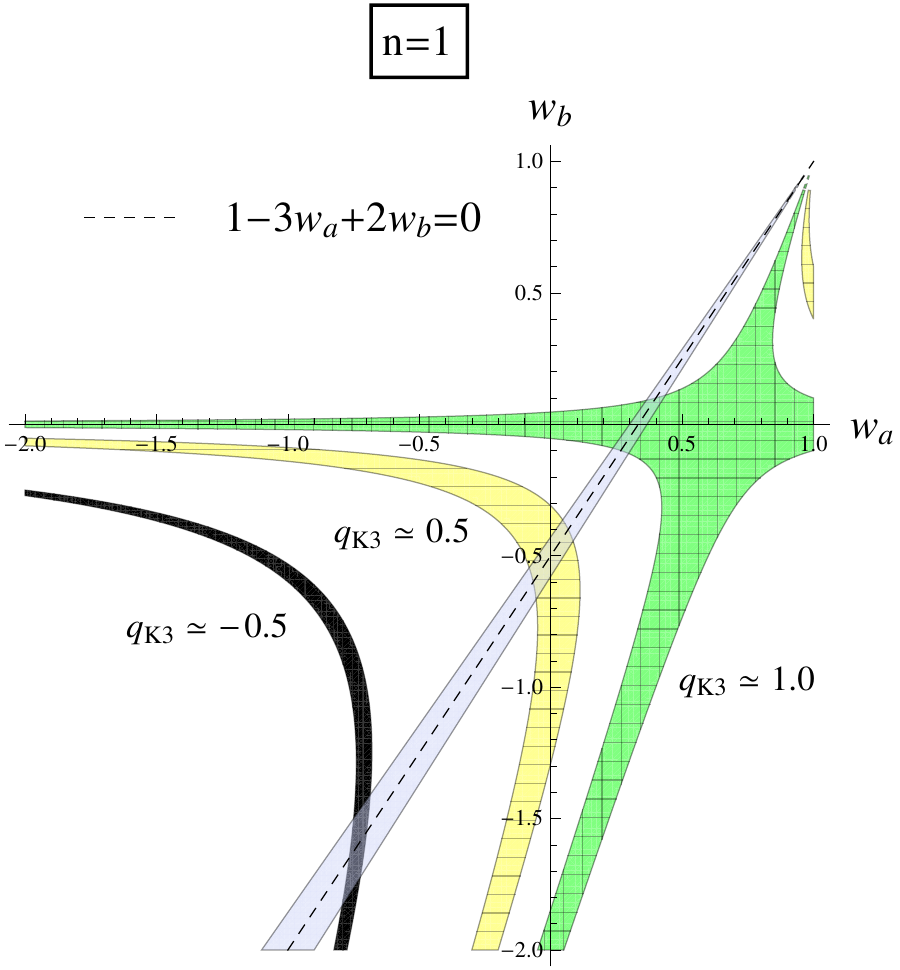}
		\caption{The triangular region (blue), around the exact stabilization condition (dashed line), shows the possible values of the w parameters that make the $K3$ solution satisfy the loose stabilization constraint \eqref{cons1} for today. The three curved regions (black, yellow - horizontal lines, green - full grid) represent possible values of these parameters that give an increasing (as we go towards $(1,1)$) value for q.}
		\label{consgraph}
\end{figure}

Constraints \eqref{stabcons} and the equivalent of \eqref{cons1} for the $K3$ solution, obtained by using \eqref{k3sol}, are depicted in Figure \ref{consgraph}, by the dashed line and the triangular region respectively. A similar treatise gives the corresponding region for constraint \eqref{cons2}.

Moreover, we can use \eqref{k3sol} to specify the regions that produce specific values for the deceleration parameter
		\[
			q=-1-\frac{\dot{H}}{H^2}~,
		\]
whose accepted value for today is $q\cong-0.6$. Various regions that correspond to different values of $q^{K3}$ are depicted in Figure \ref{consgraph}.

So what is ultimately being performed is the narrowing down of the regions of the plane of $w_a$, $w_b$ that produce specific behaviors for the Hubble parameters of the K3 solution (and thus for any other solution of the form \eqref{gensol} that converges on it). If we impose at the same time the standard cosmological evolution of	the EoS parameter $w$ to $w_a$,	(see Section \ref{specificsec}), when passing from era to era, we will have to adopt a very specific corresponding behavior for $w_b$, in order for all of the above constraints to be met.
		
One more thing that needs to be studied in terms of the EoS parameters is the decay of the perturbations of the $K3$ solution, since it ends up attracting a whole family of solutions of the form \eqref{gensol}. Setting:
		\[
		H_a(t)=H_a^{K3}(t)+H_a^{per}(t),\qquad H_b(t)=H_b^{K3}(t)+H_b^{per}(t)
		\]
		in equations \eqref{fehp} and disregarding all the non linear perturbative terms, we get the system of equations
		\begin{subequations}
\label{k3per}
\begin{eqnarray}
  \dot{H}_a^{per}&=&\frac{2 H_a(0) \bigl[\bigl(3 (w_a^2-1)+n (w_b-3 w_a^2 +3 w_a w_b-1)\bigr) H_a^{per}(t)+n w_b (w_a-n w_a+n w_b-1) H_b^{per}(t)\bigr]}{2+2 (n-1) w_a-2 n w_b+\bigl(3-3 w_a^2+n (1+3 w_a^2-6 w_a w_b+2 w_b^2)\bigr) t H_a(0)} \label{k3pera} \\ \nonumber
\\	
  \dot{H}_b^{per}&=&\frac{2 H_a(0) \bigl[3 w_a (3 w_a-2 w_b-1) H_a^{per}(t)+\bigl(3 (w_a-1)+n (3 w_a w_b-2 w_b^2-1)\bigr) H_b^{per}(t)\bigr]}{2+2 (n-1) w_a-2 n w_b+\bigl(3-3 w_a^2+n (1+3 w_a^2-6 w_a w_b+2 w_b^2)\bigr) t H_a(0)} \label{k3perb}
\end{eqnarray}
\end{subequations}
which is integrable for every n and constant $w$ parameters. However, its solution is rather cumbersome so we will only present here the behavior of the perturbations as a function of t for $n=1$, which for both $H_a^{per}$ and $H_b^{per}$ is
\[
	H_a^{per},H_b^{per} \propto t^{\frac{-4 + 3 w_a + w_b}{2 - 3 w_a w_b + w_b^2}}
\]
For all the interesting pairs of $w$ parameters (and specifically for those that guarantee a stabilized extra space), the above perturbations are decaying with time, since the exponent of t is negative in regions 2 and 3 of Figure \ref{consgraphs}, so one sees the convergence on $K3$ of all cosmologically relevant (in the sense of the $w$ parameters) solutions that are close to it.

If the same procedure were to be followed for the Kasner solutions for $n=1$ (meaning solution \eqref{kasner1n1} and $H_a=0$), it can be shown that the fluctuations evolve as
\[
	H_a^{per},H_b^{per} \propto t^{-w_b}
\]

Concluding this section, we reiterate what has been argued so far. Firstly we show that the metric \eqref{metric} has 3 particular solutions in the form of \eqref{ansatz} given by \eqref{ckas}. Moreover, its ``general" solution is \eqref{gensol}. Any solution of the form \eqref{gensol} converges on one of the particular solutions mentioned, of which only one is EoS dependent, namely $K3$. So it is interesting to see under what circumstances it can produce a viable cosmological evolution. To do that we quantify a number of constraints and impose them on the $K3$ solution, and by doing so, we have actually imposed them on any other solution of the form \eqref{gensol}, that converges on it.

This offers great freedom in terms of the initial conditions of $H_a$ and $H_b$, since the resulting cosmology depends only on the EoS parameters that specify the corresponding $K3$ solution, but at the cost of having to impose a specific, and, indeed, rather exotic from the era of matter domination and on, behavior of $w_b$. So it is stressed out that such a behavior needs to be motivated through a specific mechanism (for phantom energy scenarios for example see \cite{Caldwell:1999}, \cite{Caldwell:2003}, while for a possible motivation through string theory see \cite{Brand:1989}). Alternatively, the picture could be fundamentally different if the equation of state of the extra dimensional sector is not of the simple form $p_b=w_b \rho$.

Regardless, to demonstrate the ease that the knowledge of the corresponding $K3$ solution of such a scenario provides, we will reconstruct an evolution, very similar to that of the $\Lambda$CDM, in the next section, in the particular UED framework set by metric \eqref{metric}.

		\section{A reconstruction from today until the era of Radiation Domination} \label{specificsec}
		
		To model the desired behavior for the $w$ parameters as outlined above we will use transitions between values of $w_a$ that are consistent with the Standard Model and the observations, while the same will be done for $w_b$ demanding only that the stabilization constraints are satisfied. The results presented here, were obtained by using the density parameter $w_a$ given by the Friedmann equations:
		\begin{align} \label{wabehavior}
				w_a=\frac{2 H_a(z) H'_a(z) (1+z)-3H_a^2(z)}{3 H_a^2(z)}
		\end{align}
and using the Hubble parameter of the $\Lambda$CDM (\cite{perivol:2004}), $H_a(z)=H_a^{today}\sqrt{\Omega_M (1+z)^3+\Omega_{DE}(1+z)^{3(1+w_{DE})}}$. By choosing specific values of the $\Lambda$CDM model for the $\Omega$ parameters, the corresponding values of $w_a$ are retrieved. One can now enforce the constraint \eqref{stabcons} , and produce the evolution of $w_b$\footnote{However, other ways of transitioning between the values of $w_a$ needed to retrieve Standard Cosmology, and the corresponding values for $w_b$, should be completely viable as well (for example a generalized Chaplygin gas, \cite{Bento:2003}), as long as they follow the reasoning presented regarding the allowed values of the $w$ parameters with respect to stabilization.}.

Moreover, appropriate choices will be made to satisfy today observational results:
\[
H_0\approx 70 \frac{km/s}{Mpc}\text{~, }q_0\approx -0.6~,
\]
while simultaneously preserving the theoretical results of Standard Cosmology regarding the evolution of the scale factor in the eras of radiation and matter domination ($t^{1/2}$ and $t^{2/3}$ respectively).

To do this, we utilize our earlier result, that essentially any solution will have converged already on a $K3$ solution, that satisfies any constraint that we want to impose. Since we know analytically the behavior of any $K3$ solution, it is very easy to quantify a variety of constraints, as shown in the previous section. The combination of these, compels us to choose from a specific region if we want to achieve an apparent stabilization and $q_0\approx -0.6$ that corresponds to the observations. Finally, the transitions of the $w$ parameters for various eras, are made to also recover the generally accepted transition from a decelerating to an accelerating expansion of the Universe at redshifts $z \approx 1-2$, as well as a minute total evolution of $b(t)$, quantified by \eqref{cons2}.

Before presenting, the final results of this work, one more thing is to be noted: the observational results are generally to be matched with the effective values of the dimensionally reduced action (which typically correspond to a Gravity plus Radion-field theory), and not directly to those corresponding to the full $3+n+1$ action. Starting from the full action:
\begin{equation} \label{action1}
S\propto \int{d^{4+n}x\sqrt{-g}(R-\mathcal{L}_{matter})}
\end{equation}
with the particular metric of our setup being written in the form:
\[
g_{AB}dx^A dx^B=g^{(4)}_{\mu\nu}dx^\mu dx^\nu+b^2(t)\gamma^{(n)}_{pq}dx^p dx^q
\]
it is straightforward, (see \cite{Bring:2003-2}), to go to the aforementioned Gravity + Radion action:
\begin{equation} \label{action2}
S\propto \int{d^{4}x\sqrt{-\hat{g}}\bigl(R-\frac{1}{2}\partial_\mu \phi\partial^\mu \phi+V_{eff}(\phi)\bigr)}
\end{equation}
by means of initially integrating out the extra dimensional terms, and then performing a Weyl transformation,
\begin{equation} \label{physmetric}
\hat{g}_{\mu \nu}=b^n(t) g^{(4)}_{\mu\nu}
\end{equation}
which leads to the extra dimensions' scale factor being realized from an effective 4-D point of view as a scalar field in a potential:
\[
	\phi\propto lnb \qquad V_{eff}(\phi)=f(\mathcal{L}_{matter},\phi)
\]
Of course, the Weyl transformation changes the time and the 3-D scale factor that an effective 4-D observer would perceive as:
\begin{equation} \label{timecor}
	t_{eff}=\int b^{n/2}(t)dt +const \equiv g(t)\rightarrow \text{ }t=g^{(-1)}(t_{eff})
\end{equation}
\begin{equation} \label{scalecor}
	a_{eff}(t_{eff})=b^{n/2}\bigl(g^{(-1)}(t_{eff})\bigr) a\bigl(g^{(-1)}(t_{eff})\bigr)
\end{equation}
leading to
\begin{equation} \label{hubcor}
	H_a^{(eff)}=\Bigl[\frac{n}{2}H_b\bigl(g^{(-1)}(t_{eff})\bigr)+H_a\bigl(g^{(-1)}(t_{eff})\bigr)\Bigr]\frac{dg^{(-1)}(t_{eff})}{dt_{eff}}
\end{equation}
\begin{equation} \label{qcor}
	q_{eff}(t_{eff})=-1-\frac{dH_a^{(eff)}/dt_{eff}}{H_a^{(eff)2}}
\end{equation}

However, one can see from \eqref{timecor}-\eqref{qcor} that these corrections, with the exception of a possible scaling $b_0\approx const$, are important only if stabilization has not occurred, hence it would be necessary to take them into account only in primordial times, which however are not studied in this work.
\FloatBarrier

\begin{figure}[]
\makebox[\textwidth]{
\includegraphics[width=0.49\textwidth]{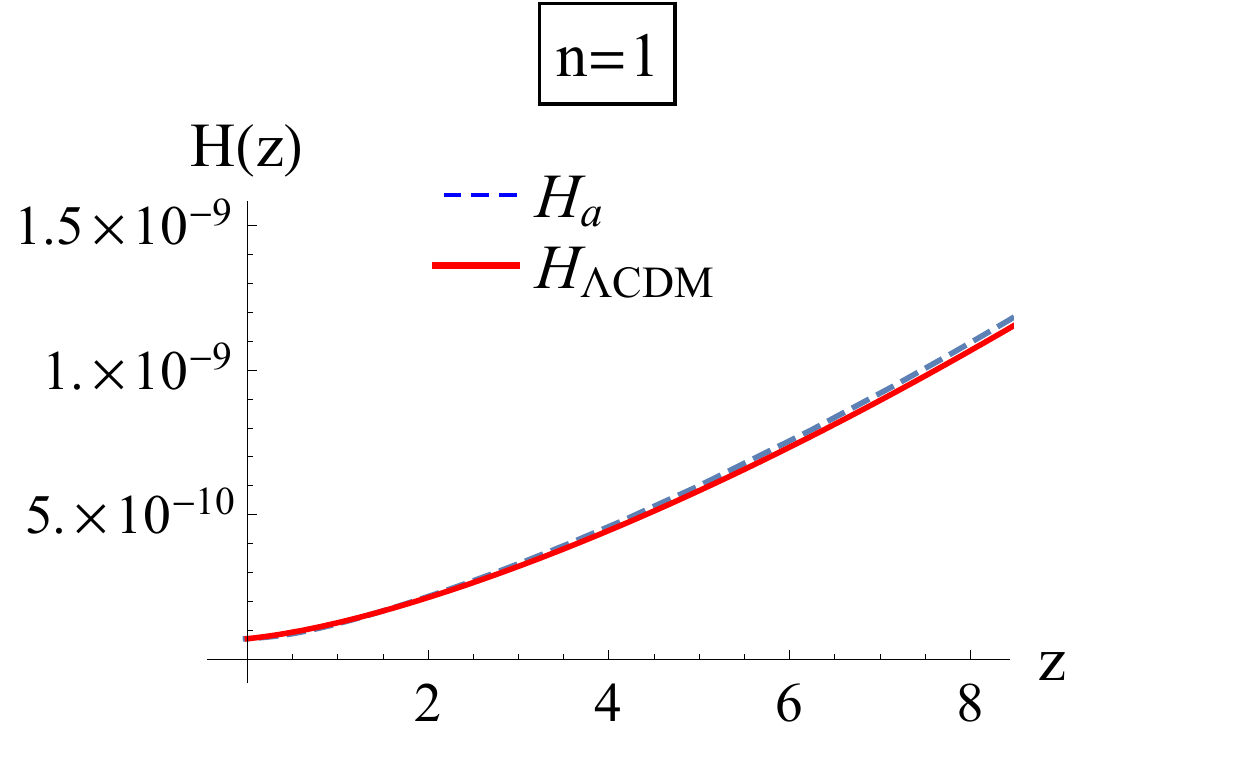}%
\hfill
\includegraphics[width=0.49\textwidth]{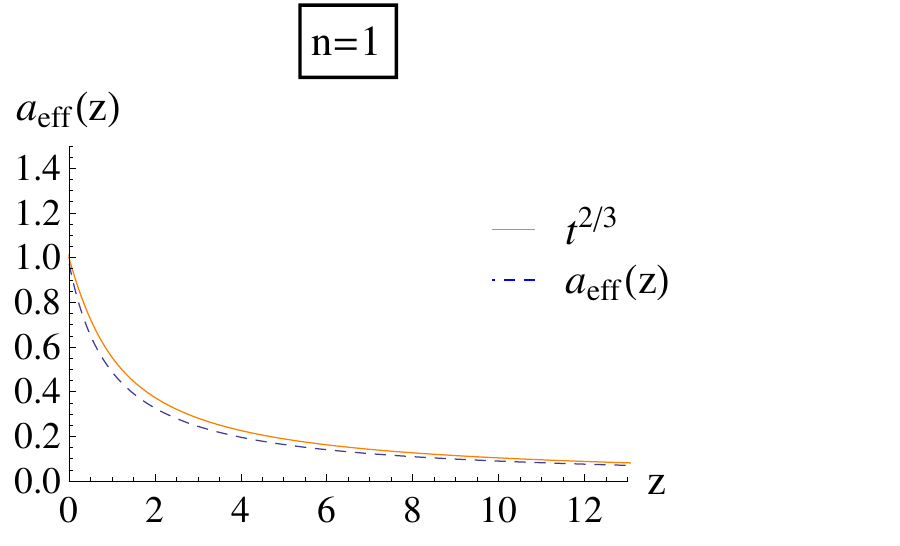}%
}\\[0.5cm]
\caption{The evolution of $H_a$ and $a(t)$ for a universe with stabilized extra space compared with their Standard Cosmology counterparts. }
\label{hubandsc}
\end{figure}

In the diagrams of Figure \ref{hubandsc} we present the evolution of the Hubble parameter $H_a$ as predicted by our model, in comparison with the evolution predicted by the $\Lambda$CDM, as well as the 3-D scale factor compared with the expected evolution for a matter dominated universe.  If we did not have an essentially stabilized $b(t)$ the evolution of the scale factor would not be the same, regardless of the choice $w_a=0$ for matter domination, since, as we see in \eqref{scalecor}, its effective value is affected by $b(t)$. The same thing is true for the radiation domination and Dark Energy era.

\begin{figure}[]
\makebox[\textwidth]{
\includegraphics[width=0.45\textwidth]{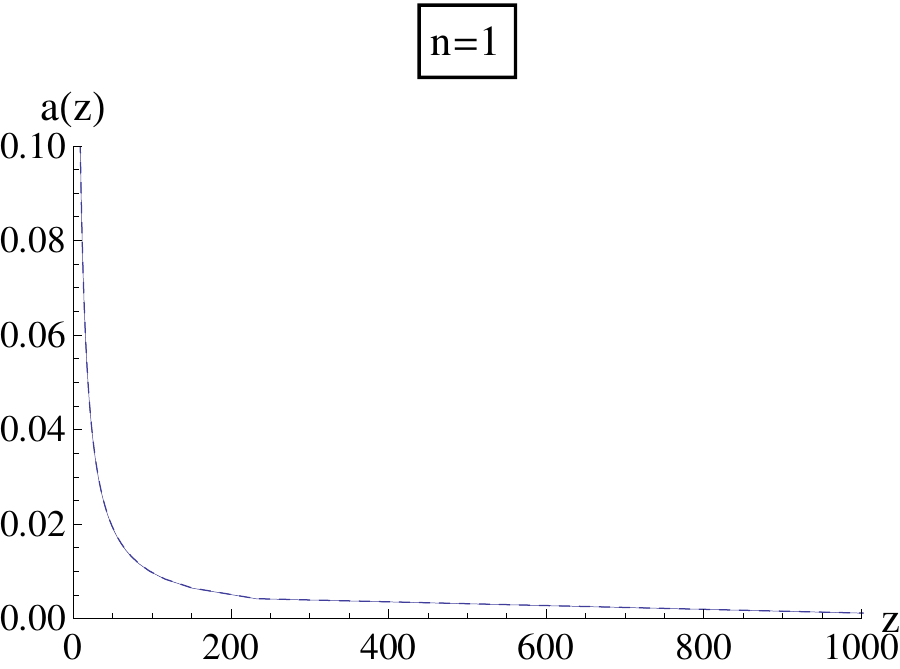}%
\hfill
\includegraphics[width=0.45\textwidth]{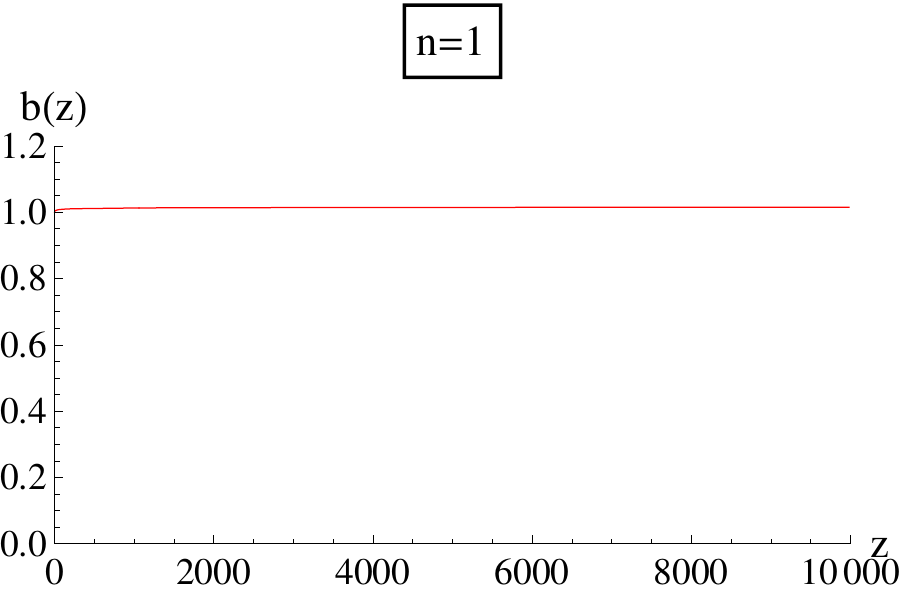}%
}\\[0.5cm]
\caption{The evolution of the scale factors of this model.}
\label{scalefactors}
\end{figure}

\begin{figure}[]
\makebox[\textwidth]{
\includegraphics[width=0.45\textwidth]{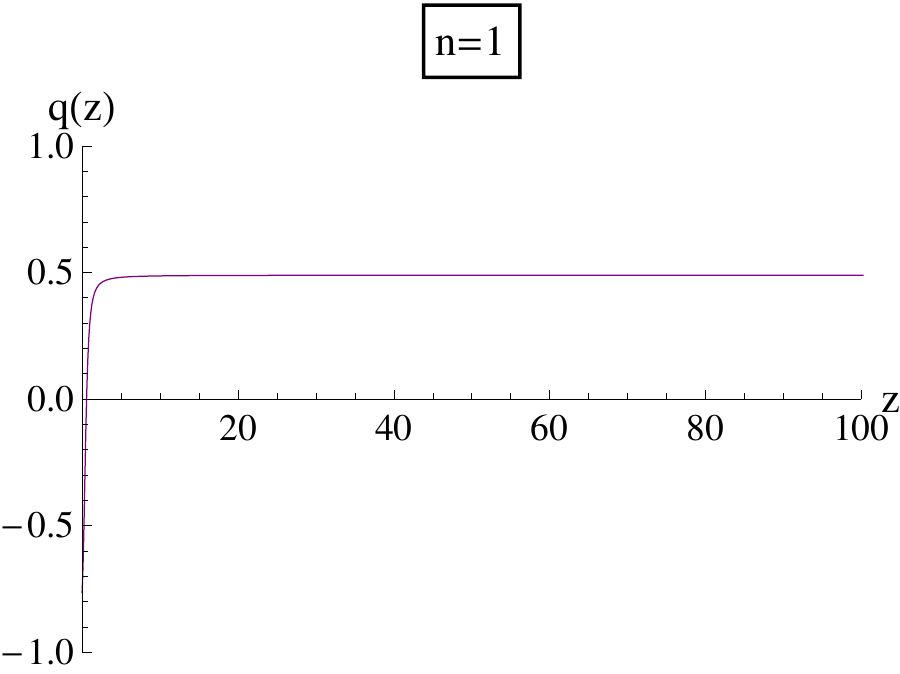}%
\hfill
\includegraphics[width=0.45\textwidth]{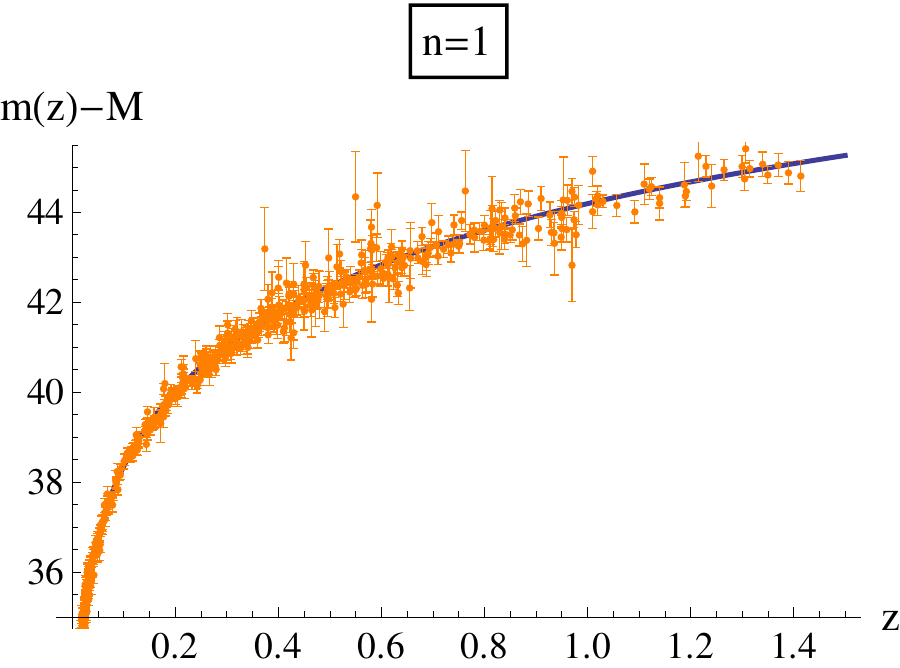}%
}\\[0.5cm]
\caption{The evolution of the deceleration parameter and a comparison of the predicted $m(z)-M$ curve with 580 SNIa points (\cite{Suzuki:2012}).}
\label{qandobs}
\end{figure}

In Figure \ref{scalefactors} we see the evolution of the scale factors, which are normalized to be $a(0)=b(0)=1$ today, and finally in Figure \ref{qandobs} we present the deceleration parameter of our model, as well as a comparison of the $m(z)-M$ curve that it predicts, with 580 SNIa observational points taken from \cite{Suzuki:2012}.

\section{K - type scale factors and their effective picture} \label{scalesec}

In this section we demonstrate the evolution of the scale factors given by the K-type solutions and their corresponding $a_{eff}(t_{eff})$, as perceived by an effective 3-D observer, which proves to be quite different from $a(t)$ when the internal space is not stabilized. From our analysis so far it is evident that the two Kasner solutions, $K1$ and $K2$, do not satisfy any stabilization condition\footnote{Though in the interesting case of an infinite dimensionality $n$, considered in \cite{Sloan}, it is shown that vacuum solutions can be stabilized.}, hence we expect a discrepancy between the evolution of $a^{eff}(t)$ and $a(t)$.

To demonstrate this, we will work in a scenario with $n=2$. By integrating \eqref{kasner1} and \eqref{kasner2} we get the scale factors for this case:

\begin{equation} \label{kasner1scale}
\left.\begin{aligned}
	a(t)&=\tilde{c_1} \bigl\lvert -\sqrt{3}+3 (\sqrt{3}-2 \sqrt{2}) H_a(0) t\bigr\rvert^{\frac{1}{-3+2 \sqrt{6}}} \\
	b(t)&=\tilde{c_2} \bigl\lvert 2 \sqrt{2}+2 \sqrt{3}+(6 \sqrt{2} +2 \sqrt{3})H_a(0) t\bigr\rvert^{-\frac{2}{2+2 \sqrt{6}}}
\end{aligned}
\right\}
\qquad \text{K1}
\end{equation}

\begin{equation} \label{kasner2scale}
\left.\begin{aligned}
	a(t)&=\tilde{c_1} \bigl\lvert -\sqrt{3}+3 (\sqrt{3}+2 \sqrt{2}) H_a(0) t\bigr\rvert^{\frac{1}{-3-2 \sqrt{6}}} \\
	b(t)&=\tilde{c_2} \bigl\lvert -2 \sqrt{2}+2 \sqrt{3}+(2 \sqrt{3}-6 \sqrt{2})H_a(0) t\bigr\rvert^{-\frac{2}{2-2 \sqrt{6}}}
\end{aligned}
\right\}
\qquad \text{K2}
\end{equation}

where $\tilde{c_1}$, $\tilde{c_2}$ are integration constants. According to \eqref{timecor} and \eqref{scalecor} we can, in principle, get the function $t(t_{eff})$ and subsequently $a_{eff}(t_{eff})$. However, one can see that it is not a trivial task, even though we have the explicit forms of $a(t)$, $b(t)$, so instead we continue qualitatively.

 For the $K1$ solution, from \eqref{kasner1scale} we see that for $t\gg t_{sing}$, $b(t)\propto t^{-\frac{2}{2+2 \sqrt{6}}}$, so from \eqref{timecor}, we deduce that in this case $t\propto t_{eff}^{\frac{6-\sqrt{6}}{5}} \approx t_{eff}^{7/10}$. So by \eqref{scalecor} we see that for $t\gg t_{sing}$, $a_{eff}\propto t_{eff}^{1/3}$, as opposed to $a\propto t^{1/2}$.

On the other hand, very close to the singularity the dimensionally reduced metric is not necessarily the physical one. Still, if we were to naively look for an ``inflationary"-like evolution of the $K1$ solution, we would conclude that neither $a(t)$, nor its effective counterpart have a fast enough evolution.

The same reasoning can be followed for the $K2$ solution, giving similar results, however one needs to take into account that in this case the singularity is in the future and not in the past.

The situation, however, can be quite different in the case of $K3$, as we will show in the following example for $n=2$. The scale factors in this case are:

\begin{equation} \label{kasner3scale}
\left.\begin{aligned}
	a(t)&=\tilde{c_1} \bigl\lvert 2 (1+w_a-2 w_b)+(5+3 w_a^2-12 w_a w_b+4 w_b^2) H_a(0) t \bigr\rvert^{\frac{2+2 w_a-4 w_b}{5+3 w_a^2-12 w_a w_b+4 w_b^2}} \\
	b(t)&=\tilde{c_2} \bigl\lvert 2 (1+w_a-2 w_b)+(5+3 w_a^2-12 w_a w_b+4 w_b^2) H_a(0) t \bigr\rvert^{\frac{2-6 w_a+4 w_b}{5+3 w_a^2-12 w_a w_b+4 w_b^2}}
\end{aligned}
\right\}
\qquad \text{K3}
\end{equation}

One immediately sees that there exist suitable values for the $w$ parameters, namely $w_a=-1$, $w_b=-2$, that make the denominators of both the exponents, as well as the numerator of b(t), go to zero. The exponent of $a(t)$ is positive in regions 2 and 3 of Figure \ref{consgraphs}, while the exponent of $b(t)$ is only positive in the subset of region 2 defined to the left of the dashed line representing the exact stabilization condition $1-3 w_a+2 w_b=0$. Hence, given a properly selected approach to the above values of the $w$ parameters, we can achieve a large positive value for the exponent of $a(t)$, and at the same time a small negative value for the exponent of $b(t)$. However, both the stability of the solution and the value of the exponent of $b(t)$ depend on the way that the aforementioned values of the $w$'s are approached. These values lie exactly on the boundary of regions 1 and 2 of Figure \ref{consgraphs} (this is true for the respective diagrams for every $n$), meaning that a fluctuation of the $w$'s from one region to the other, changes drastically the behavior of the solution. To illustrate this, we refer the reader to the top two diagrams of Figure \ref{fig:flowfigs} of the Appendix, where the different behaviors of the phase curves for values in regions 1 and 2 of Figure \ref{consgraphs} respectively, can be seen.

\section{Conclusions} \label{conclusion}

In this work we have presented a concise view of how flat, homogeneous Universal Extra Dimensions affect standard cosmological evolution, and how one can recreate a picture similar to that of the $\Lambda$CDM. In the framework of UED, this can only be done if the extra dimensions are stabilized from a very early era, since a significant fluctuation in the values of fundamental coupling constants would be measurable by experiments, or observable through deviations from models of high redshift events (like BBN).

We have managed to do so for non-exactly static, but still slowly-enough evolving extra dimensions, by using a special case solution for constant EoS parameters. This Kasner-type, particular solution actually acts as an attractor for a plethora of possible cosmologically relevant (i.e. expanding 3-D, contracting extra space) initial conditions of the Hubble parameters $H_a$, $H_b$. It is an exact solution of the Friedmann equations, and because it is analytically known, through its dependence on the EoS parameters, it can give the desired cosmological evolution compatible with the phenomenology. Then, we show that a large variety of random initial conditions yield solutions that converge rapidly on this phenomenologically correct attractor-solution. To achieve the stabilization of the extra space without using any explicit mechanism, we have to allow for a wider range of EoS parameters for both the usual and the extra spatial fluid. However, the EoS parameters need to follow a specific, codependent evolution, quantified through a very simple relation. The range of the EoS parameters is expected to arise from string theory, since it is known that strings wound around compactified dimensions give similar effects. However, a justification for a constraint connecting the EoS parameters in such a specific way, that is indeed dependent only on the dimensionality of the usual space, has yet to emerge.

Finally, we have studied how the scale factors corresponding to this special Kasner-type solution would behave in times close to its singularity. We have seen that there exists a pair of EoS parameters that can at the same time produce a very fast expanding evolution of the usual 3-D scale factor and a comparatively very slow contracting evolution of the extra spatial scale factor. However, these EoS parameters lay on the border of two regions that in general produce very different evolution patterns, meaning that a small fluctuation in their values could trigger a significant change in the evolution of the usual and extra space.

\appendix

\section{ }

We present here for reference, two flow diagrams for $n=1$ for two specific examples regarding the EoS parameters. These examples illustrate the behavior of the general solution, without trying a priori to match anything with realistic cosmological results. As stated in section \ref{solutionsec}, the evolution of a system of this type depends on the values of the $w$ parameters and, of course, on the position of the initial values $H_a^{(i)}$, $H_b^{(i)}$ with regard to the curves of the Kasner type curves $K1$, $K2$, $K3$ in the phase space. Since $K1$ and $K2$ (which for $n=1$ reduces to the axis $H_a=0$) have constant ratios, the value of the $K3$ solution's ratio is the deciding factor as to where its curve is with regard to the $K1$ and $K2$ curves. That, in turn, in combination with the initial values and the signs of the $r.h.s$ factors of \eqref{fehp}, decides the attractor-curve in each case. These behaviors are qualitatively the same for any $n$.

\FloatBarrier

\begin{figure}[h]
\makebox[\textwidth]{
\includegraphics[width=0.42\textwidth]{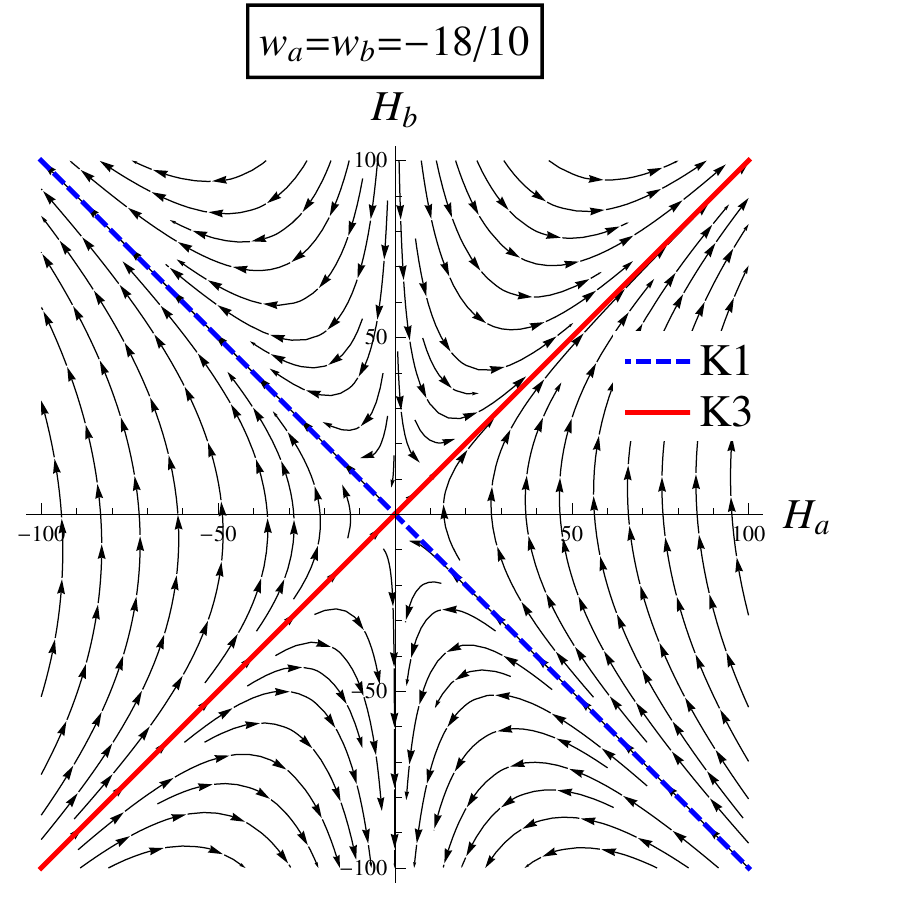}%
\hfill
\includegraphics[width=0.42\textwidth]{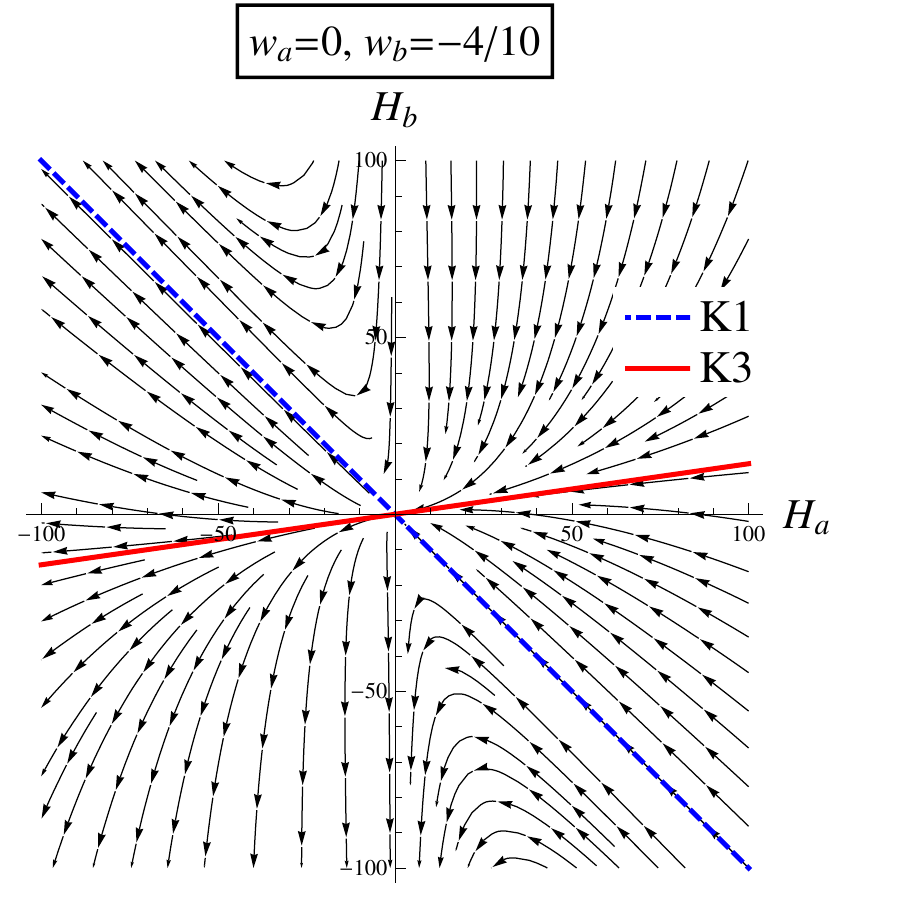}%
}\\[0.5cm]
\caption{Two characteristic examples for $w$ parameters in regions 1 and 2 of Figure \ref{consgraphs} respectively. Qualitatively these behaviors remain the same for any $n$.}
\label{fig:flowfigs}
\end{figure}

\FloatBarrier
\clearpage


\end{document}